\pdfoutput=1
\documentclass[12pt]{article}

 \usepackage{booktabs}
\usepackage{jheppub}
\usepackage[a4paper,left=5cm,right=0.0cm,top=6cm,bottom=1cm]{geometry}

\makeatletter
\renewcommand{\affiliation}[2][]{%
  \affiltrue
  \if!#1!%
    \affil@toks=\expandafter{\the\affil@toks{\item[]{\footnotesize #2}}}%
  \else
    \affil@toks=\expandafter{\the\affil@toks{\item[$^{#1}$]{\footnotesize #2}}}%
  \fi
}
\makeatother

\usepackage{graphicx}
\usepackage{subcaption}
\usepackage{amssymb}
\usepackage{amsmath}
\usepackage{amsfonts}
\usepackage{multirow}
\usepackage{color}
\usepackage{xcolor}
\usepackage{url}
\usepackage{ulem}
\usepackage{float}

\usepackage{pifont}

\usepackage{soul}
\usepackage[english]{babel}
\usepackage[T1]{fontenc}
\usepackage{lmodern}
\usepackage[utf8]{inputenc}  
\usepackage{bbm}

\usepackage{cancel}
\newcommand{\met}{\cancel{E}_{T}}
\def\nicefrac#1#2{\hbox{$\frac{#1}{#2}$}}

\newcommand{\be}{\begin{equation}}
\newcommand{\ee}{\end{equation}}
\newcommand{\bea}{\begin{eqnarray}}
\newcommand{\eea}{\end{eqnarray}}
\newcommand{\Z}{\mathbb{Z}}
\newcommand{\GeV}{{\ensuremath\rm \; GeV}}
\newcommand{\calchep}{\texttt{CalcHEP}}
\newcommand{\madgraph}{\texttt{MadGraph5\_aMC@NLO}}
\newcommand{\pythia}{\texttt{Pythia}}
\newcommand{\checkmate}{\texttt{CheckMATE}}
\usepackage{hyperref}
\hypersetup{
  colorlinks   = true, 
  urlcolor     = blue, 
  linkcolor    = blue, 
  citecolor   = blue 
}

\graphicspath{{plots/}}

\newcommand\ReDiag{\mathop{%
  \raise .5pt\hbox{[}%
  \widetilde{\mathrm{Re}}%
  \raise .5pt\hbox{]}}}
\newcommand\ReOffDiag{\mathop{%
  \raise .5pt\hbox{$\llbracket$}%
  \widetilde{\mathrm{Re}}%
  \raise .5pt\hbox{$\rrbracket$}}}

\newcommand{\gev}{\,\, \mathrm{GeV}}

\newcommand\CM{\texttt{CheckMATE}}

\def\gmin2{\ensuremath{(g-2)_\mu}}

\def \met  {\mbox{${E\!\!\!\!/_T}$}}

\definecolor{Orange}{named}{orange}
\definecolor{Purple}{named}{purple}
\definecolor{Lightblue}{cmyk}{0.9,0.1,0.1,0.3}
\definecolor{dgelborange}{cmyk}{0.,0.3,0.5, 0.}
\definecolor{Lila}{rgb}{0.5,0.,1}
\definecolor{Darkgreen}{rgb}{0.,.7,0.2}

\allowdisplaybreaks
\definecolor{darkgreen}{rgb}{0.0, 0.4, 0.0} 
\definecolor{darkorange}{rgb}{1.0, 0.4, 0.4}

\usepackage{xcolor}

\makeatletter
\renewcommand{\@email}[1]{\href{mailto:#1}{\ttfamily\footnotesize #1}}
\makeatother
\begin{document} 
\title{\hfill~\\[-20mm]
\begin{footnotesize}
\hspace{12.0cm}\normalfont{DIAS-STP-26-14}\\
\hspace{12.0cm}\normalfont{CERN-TH-2026-145}\\
\hspace{12.0cm}\normalfont{RBI-ThPhys-2026-21}\\
\end{footnotesize}
\vspace{7mm}
\boldmath
\begin{center}
Soft-Dimuon Signature from
Two-Component Scalar Dark Matter
at the LHC
\end{center}
}
\author[\,1,2]{A.~Belyaev,}
\author[\,1]{M.~Chakraborti,}
\author[\,1]{S.~Chen,}
\author[\,3]{A.~Dey,}
\author[\,4,5,6]{V.~Keus,}
\author[\,7]{R.~Mahbubani,}
\author[\,1,2,3]{S.~Moretti}

\affiliation[1]{School of Physics and Astronomy, University of Southampton, SO17 1BJ Southampton, United Kingdom}
\affiliation[2]{Particle Physics Department, Rutherford Appleton Laboratory, Chilton, Didcot OX11 0QX, United~Kingdom}
\affiliation[3]{Department of Physics and Astronomy, Uppsala University, Box 516, SE-751 20 Uppsala, Sweden}
\affiliation[4]{Theoretical Physics Department, CERN, 1 Esplanade des Particules, Geneva 23, CH-1211, \mbox{Switzerland}}
\affiliation[5]{School of Theoretical Physics, Dublin Institute for Advanced Studies, 10 Burlington Road, Dublin, D04 C932, Ireland}
\affiliation[6]{Department of Physics and Helsinki Institute of Physics, Gustaf Hallstromin katu 2, FIN-00014, University of Helsinki, Finland}
\affiliation[7]{Rudjer Boskovic Institute, Division of Theoretical Physics, Bijenicka 54, HR-10000 Zagreb, Croatia}

\emailAdd{\text{a.belyaev@soton.ac.uk, m.chakraborti@soton.ac.uk, shu.chen@soton.ac.uk}}
\emailAdd{\text{atri.dey@physics.uu.se, venus@stp.dias.ie, rakhi@irb.hr, stefano.moretti@cern.ch}}
\vspace*{-1.50truecm}

\abstract{\footnotesize
In this work we explore the potential of the Large Hadron Collider (LHC) to probe a two-component
scalar Dark Matter (DM) scenario in the opposite-sign dimuon ($\mu^+\mu^-$) plus missing transverse energy ($\met$) final state, { accompanied by a hard jet ($j$). In particular, the signal is characterised by a soft dimuon system with an invariant mass well below $m_Z$.}
We work within the framework of a 3-Higgs Doublet Model (3HDM) containing
one active Higgs doublet and two inert scalar doublets, hence termed the I(2+1)HDM.
{ A $\Z_2\times \Z_2'$ symmetry stabilises the lightest neutral scalar from each inert doublet, giving rise to two scalar DM candidates.}
We provide a simple mapping of the model parameter space in terms of the
masses of the two DM candidates and the mass differences between {each DM candidate and the corresponding next-to-lightest scalar state}.
{We perform a detector-level Monte Carlo analysis and devise a dedicated cut-based selection, including a transverse-mass requirement tailored to the signal topology. For our benchmark, we obtain $S/B\simeq 9.8\%$ and a statistical-only significance $S/\sqrt{B}=1.35$ at the LHC Run 3 with ${\cal L}=300~{\rm fb}^{-1}$ while a statistical-only extrapolation to ${\cal L}=4~{\rm ab}^{-1}$ gives $S/\sqrt{B}=4.93$.}
The two dark sectors produce a characteristic double-bump structure in the dimuon invariant-mass distribution before the full selection. However, this feature is not retained with sufficient significance after the cuts optimised for inclusive signal sensitivity are enforced, making it difficult to establish experimentally that the signal originates from two DM components. The benchmark is underabundant and is therefore interpreted as a subdominant two-component DM scenario, yet,  the collider analysis itself is independent of this cosmological behaviour.
Finally, although the numerical analysis is performed within the I(2+1)HDM for illustrative purposes,
{ the results apply more broadly to weakly interacting sectors with analogous electroweak associated production and cascade decays, in which a heavier state is separated from the DM candidate by less than $m_Z$ yields a soft muon pair through an off-shell $Z$ boson.}
}

\maketitle

\section{Introduction}
\label{sec:intro}

Despite the remarkable success of the Standard Model (SM) in explaining
most of the observed experimental data, {  the nature of Dark Matter (DM) remains unknown. Explaining DM within particle physics requires extending the SM with new particles and interactions. Among the best-studied possibilities are Weakly Interacting Massive Particles (WIMPs), which can produce observable signatures in Direct Detection (DD) and Indirect Detection (ID) experiments as well as at particle colliders.}
Thus, the detection of WIMPs has been one of the primary motivations for the Large Hadron Collider (LHC) and
the anticipated future collider experiments.

{ In the vast literature on WIMPs, significant effort has been devoted to constructing minimal dark-sector scenarios in which a single DM species accounts for the entire observed relic density of the Universe. However, the rich variety of particles and interactions present in the SM itself naturally motivates considering non-minimal dark sectors containing more than one stable particle. Consequently, the exploration of multi-component WIMP DM has gained substantial momentum in recent years. Multi-component scenarios can give rise to collider signatures that differ qualitatively from those of single-component DM, since several dark states with different masses and mass splittings may contribute simultaneously to the same final state. In this work, we explore the collider phenomenology of a dark sector featuring two scalar DM candidates.}

As a typical example of a two-component scalar DM scenario, we choose to work within
the framework of a specific 3-Higgs Doublet Model (3HDM), with one active Higgs doublet and two inert scalar doublets,
also referred to in the literature
as the I(2+1)HDM~\cite{Keus:2014jha,Keus:2014isa,Keus:2015xya,Cordero-Cid:2016krd,Cordero:2017owj,Cordero-Cid:2018man,Keus:2019szx,Cordero-Cid:2020yba,Hernandez-Sanchez:2020aop,Hernandez-Sanchez:2022dnn,Dey:2024epo}. The Inert Doublet Model (IDM), also known as the I(1+1)HDM,
with one active Higgs doublet and one inert scalar doublet, has been studied extensively in the literature~\cite{Deshpande:1977rw}
in the context of single-component DM. To ensure the stability of the DM, a $\Z_2$ symmetry is typically imposed
on such a model under which only the inert doublet is considered to be odd.
Therefore, the lightest $\Z_2$-odd neutral particle stemming from the inert doublet can
serve as a viable candidate for DM. The I(2+1)HDM can be considered as the simplest extension of the IDM required to accommodate a two-component scalar DM scenario. In this case, the discrete
$\Z_2$ symmetry is extended to a $\Z_2\times \Z_2^\prime$ one where the first (second) inert doublet
is taken to be odd under the $\Z_2~(\Z_2^\prime)$ symmetry. This opens up the possibility
of two scalar DM candidates which turn out to be the lightest neutral scalar
from each of the inert doublets. The phenomenology of such a scenario
in DD and ID experiments searching for DM has been studied previously in
Ref.~\cite{Hernandez-Sanchez:2020aop}, while for the case of LHC experiments this has been done in Refs.~\cite{Hernandez-Sanchez:2022dnn,Dey:2024epo}.
In particular, Refs.~\cite{Hernandez-Sanchez:2022dnn,Dey:2024epo} have pointed out that the presence
of two DM candidates leads to specific shapes in the distributions of various observables
at the LHC. { However, a detector-level analysis including the relevant SM backgrounds and a dedicated event selection has not previously been performed for the soft-dimuon signature considered here.}

\begin{figure}[t!]
\centering
\includegraphics[scale=0.4]{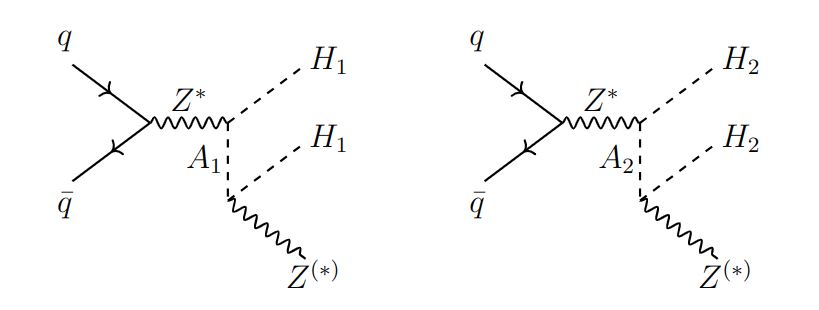}
\caption{Diagrams leading to the $\mu^+\mu^-+\met$ final state mediated by { an off-shell} $Z$ boson. { The additional hard ISR jet required in the analysis is not shown.}}
\label{fig:feyn}
\end{figure}

The LHC signal that we focus on in this work contains two visible muons\footnote{We use muons in preference to electrons { because soft muons can be reconstructed with low transverse-momentum, $p_T$, thresholds and are less affected by fake and non-prompt backgrounds in the relevant phase-space region.}} and
missing transverse energy, $\met$, in the final state generated by the production
of the next-to-lightest particle in the dark sector and { its} cascade decay into
the DM candidate for each inert doublet. Such a signal has been studied previously in the context of the
I(1+1)HDM in Ref.~\cite{Belyaev:2022wrn}. Operationally, in order to increase the signal sensitivity, we { require the presence of a hard jet from Initial State Radiation (ISR), against which the dark-sector system recoils, thereby enhancing $\met$.} The schematic diagram for the process at the LHC, { with the ISR jet omitted for simplicity,} is shown in Fig.~\ref{fig:feyn},
where $A_i$ and $H_i$ with $i=1,2$ denote the next-to-lightest particle and the DM candidate from
the two inert sectors, respectively.

{ The signal kinematics in each sector are primarily controlled by two quantities: the DM mass, $M_{H_i}$, and the mass splitting $\Delta M_i=M_{A_i}-M_{H_i}$. Since the $ZA_iH_i$ coupling is fixed by the electroweak gauge interaction, these masses determine the production kinematics and the endpoint of the soft-dimuon invariant-mass distribution. The simultaneous contribution of two sectors with different values of $\Delta M_i$ can therefore generate a characteristic double structure in the dimuon spectrum.}

Such a simple parametrisation of the model parameter space in the context of
the $\mu^+\mu^-+\met+j$ signal makes our analysis { applicable beyond the specific I(2+1)HDM benchmark considered here, in particular to weakly interacting sectors with analogous electroweak associated production and cascade decays through an off-shell $Z$ boson.}

The plan of the paper is as follows.
In Sec.~\ref{sec:model} we briefly describe the I(2+1)HDM. In Sec.~\ref{sec:cons} we illustrate
the main theoretical and experimental constraints that have led to
the selection of the Benchmark Point (BP) considered in this analysis. We detail the
strategy for our collider analysis for such a BP in
Sec.~\ref{sec:ana}. Following a discussion of the relevant backgrounds in Sec.~\ref{sec:bg},
our main results are presented in Sec.~\ref{sec:results}, where we show
the distributions of the relevant collider observables, explain the reason behind the cuts employed
to suppress background events and finally illustrate the outcome of our cut-and-count analysis. We also
present the results of a generic scan of the model parameter space in this part, showing
the signal cross section, efficiency and significance at the LHC.
Finally, we summarise in Sec.~\ref{sec:summary}.


\section{The Model}
\label{sec:model}
In this section we describe the scalar sector of the I(2+1)HDM. 
The most general scalar potential of a 3HDM with $\Z_2\times \Z_2^\prime$
symmetry can be written  as~\cite{Ivanov:2011ae,Keus:2013hya,Hernandez-Sanchez:2022dnn,Dey:2024epo},
\begin{eqnarray}
\label{eq:pot}
V_{\Z_2 \times \Z_2'} &=& - \mu^2_{1} (\phi_1^\dagger \phi_1) -\mu^2_2 (\phi_2^\dagger \phi_2) - \mu^2_3(\phi_3^\dagger \phi_3) \nonumber
+ \lambda_{11} (\phi_1^\dagger \phi_1)^2+ \lambda_{22} (\phi_2^\dagger \phi_2)^2  + \lambda_{33} (\phi_3^\dagger \phi_3)^2 \nonumber\\
&& + \lambda_{12}  (\phi_1^\dagger \phi_1)(\phi_2^\dagger \phi_2)
 + \lambda_{23}  (\phi_2^\dagger \phi_2)(\phi_3^\dagger \phi_3) + \lambda_{31} (\phi_3^\dagger \phi_3)(\phi_1^\dagger \phi_1) \nonumber\\[0.5mm]
&& + \lambda'_{12} (\phi_1^\dagger \phi_2)(\phi_2^\dagger \phi_1)
 + \lambda'_{23} (\phi_2^\dagger \phi_3)(\phi_3^\dagger \phi_2) + \lambda'_{31} (\phi_3^\dagger \phi_1)(\phi_1^\dagger \phi_3)  \nonumber\\[1mm]
&&  +\lambda_1 (\phi_1^\dagger \phi_2)^2 + \lambda_2(\phi_2^\dagger \phi_3)^2 + \lambda_3 (\phi_3^\dagger \phi_1)^2 + \mathrm{h.c.}, \nonumber
\end{eqnarray}
where the terms in the first three lines are invariant under any
general phase rotation while the terms in the last line make explicit
the underlying $\Z_2\times \Z_2^\prime$ symmetry of the model. The $\Z_2\times \Z_2^\prime$
charge assignment is such that the doublet $\phi_3$ as well as all the SM fields are even under
both $\Z_2$ and $\Z_2^\prime$ whereas only the doublet $\phi_1 (\phi_2)$ is odd under
$\Z_2 (\Z_2^\prime)$. This charge assignment readily implies a Type-I Yukawa 
interaction structure, with only the doublet $\phi_3$ coupling to the SM fermions. This eliminates
the possibility of any tree-level Flavour Changing Neutral Current (FCNC) in this model.
{ We work in a CP-conserving scenario and take all parameters in the scalar potential to be real.}

We are interested in the  configuration of the model where only
one of the doublets acquires a Vacuum Expectation Value (VEV), leaving the other 
two doublets inert. The vacuum state can be expressed as
\begin{eqnarray}
\label{eq:vac}
\phi_1= \left(\begin{array}{c}
\scriptstyle H^+_1  \\
{\frac{H_1+iA_1}{\sqrt{2}}} \\
\end{array}\right) \,, \qquad
\phi_2= \left(\begin{array}{c}
\scriptstyle H^+_2 \\
{\frac{H_2+iA_2}{\sqrt{2}}} \\
\end{array}\right) \,, \qquad
\phi_3= \left(\begin{array}{c}
\scriptstyle G^+ \\
{\frac{v + h +iG^0}{\sqrt{2}}} \\
\end{array}\right) \,, 
\end{eqnarray}
where { $H_i$ ($A_i$) denotes the CP-even (CP-odd) inert neutral scalar} and
$H_i^+$ denotes the charged states, 
with $i=1,2$. Here, the $\phi_3$ field with a non-zero VEV $v$ serves as the SM Higgs boson doublet with 
the ensuing physical state $h$  coupling to the SM fermions and gauge bosons with strengths identical
to those of the SM Higgs boson. The states $G^+$ and $G^0$ become the
Goldstone bosons after Electro-Weak Symmetry Breaking (EWSB) has occurred.

The mass of the SM Higgs boson comes out to be
\begin{equation}
\label{eq:SM-higgs}
m^2_h = 2\mu_3^2 = 2 v^2 \lambda_{33}
\end{equation}
while those of the inert physical states can be written as~\cite{Hernandez-Sanchez:2022dnn,Dey:2024epo}
\begin{eqnarray}
\label{eq:mass}
m^2_{H_1} &=&  -\mu_1^2 + \Lambda_3 v^2, \qquad\, m^2_{H_2} = -\mu_2^2 + \Lambda_2 v^2, \\
m^2_{A_1} &=& -\mu_1^2 + {\Lambda^\prime}_3 v^2, \qquad\, m^2_{A_2}  = -\mu_2^2 + {\Lambda^\prime}_2 v^2, \\
m^2_{H^\pm_1} &=& -\mu^2_1 +\frac{1}{2}\lambda_{31}v^2, \quad\, m^2_{H^\pm_2} = -\mu^2_2 +\frac{1}{2}\lambda_{23} v^2,
\end{eqnarray}
 with the $\Lambda$'s defined as
 {
 \begin{eqnarray}
\Lambda_3 &=&\frac{1}{2}(\lambda_{31}+\lambda'_{31} +2\lambda_3 ), \qquad\,
\Lambda_3^\prime =  \frac{1}{2}(\lambda_{31}+\lambda'_{31} -2\lambda_3), 
\\
\Lambda_2 &=& \frac{1}{2}(\lambda_{23}+\lambda'_{23} +2\lambda_2), \qquad\,
\Lambda_2^\prime = \frac{1}{2}(\lambda_{23}+\lambda'_{23} -2\lambda_2).
 \end{eqnarray}
 }

In principle, any particle in the dark sector $(H_i, A_i, H^\pm_i)$ can be the lightest. We dismiss the possibility of $H^\pm_i$ being the lightest, as it would mean that the DM candidate is a charged particle. Choosing between $H_1$ and $A_1$ (or $H_2$ and $A_2$) is related only to a change of the sign of the quartic parameter $\lambda_{3}$($\lambda_2$) and has no impact on the phenomenology. 
Therefore, by demanding that $m_{H_i} < m_{A_i}, m_{H^\pm_i}$, we obtain the following relations between parameters:
\be
\lambda_2<0, \;\;\; \lambda_3<0, \;\;\; \lambda_{31}'+ 2\lambda_3<0, \;\;\;  \lambda_{23}'+ 2\lambda_2<0. \label{Hlightest}
\ee
{ Notice that, unlike in many $\Z_N$-symmetric models (e.g., see \cite{Aranda:2019vda,Arroyo-Urena:2026vgf} for the case $N=3$), the lightest state from each inert doublet is automatically stable, independently of their relative masses, since the two states are stabilised by different $\Z_2$ symmetries.}

The parameters of the potential can be rewritten in terms of physical observables, such as masses and couplings. For our study, { we choose the following masses and couplings as the physical parameter basis:}
\be
v,\, m_h,\, m_{H_1}, \,m_{H_2},\, m_{A_1},\, m_{A_2},\, m_{H^\pm_1},\, m_{H^\pm_2},\, \Lambda_2,\, \Lambda_3,\, \Lambda_1,\, \lambda_{11},\, \lambda_{22},\, \lambda_{12}, \,\lambda'_{12} \label{physpar},
\ee
where the $\Lambda_i$ couplings are 
\be 
g_{hH_1H_1} = 2 \Lambda_3 \,, \qquad g_{hH_2H_2} = 2 \Lambda_2 \,,
\qquad g_{H_1H_1H_2H_2} = 2 \Lambda_1  \,,
\ee
as they appear in the interaction terms in the Lagrangian
\be 
-\mathcal{L} \, \supset \, \nicefrac{v}{2}\, g_{hH_1H_1}\, h H_1 H_1 
\,+\, \nicefrac{v}{2} \,g_{hH_2H_2} \,h H_2 H_2 
\,+ \, \nicefrac{1}{4} \, g_{H_1H_1H_2H_2} \,H_1 H_1 H_2 H_2 \,.
\ee
The self-couplings $\lambda_{11},\, \lambda_{22},\, \lambda_{12},\, \lambda'_{12}$ correspond exactly to the terms in Eq.~\eqref{eq:pot}, while { the remaining parameters are related to our chosen physical basis as follows:}
\bea
&& \mu_1^2=-m_{H_1}^2+\Lambda_{3}v^2 \,, \\[0.5mm]
&& \lambda_{3}=(m_{H_1}^2-m_{A_1}^2)/(2v^2)\,, \\[0.5mm]
&& \lambda_{31}'=(m_{H_1}^2+m_{A_1}^2-2m_{H^\pm_1}^2)/v^2\,, \\[0.5mm]
&& \lambda_{31}=2\Lambda_{3}-2\lambda_{3}-\lambda_{31}'\,, \\[0.5mm]
&& \mu_2^2=-m_{H_2}^2+\Lambda_{2}v^2\,, \\[0.5mm]
&& \lambda_{2}=(m_{H_2}^2-m_{A_2}^2)/(2v^2)\,,\\[0.5mm]
&& \lambda_{23}'=(m_{H_2}^2+m_{A_2}^2-2m_{H^\pm_2}^2)/v^2\,, \\[0.5mm]
&& \lambda_{23}=2\Lambda_{2}-2\lambda_{2}-\lambda_{23}'\,,\\[0.5mm]
&& \lambda_1 = \Lambda_1 - (\lambda_{12}+\lambda'_{12})/2\,.
\eea

\section{Theoretical and Experimental Constraints}
\label{sec:cons}

The parameters of the potential $V_{\Z_2 \times \Z_2'}$ are subject to a number of theoretical and experimental constraints. { Below, we summarise the constraints imposed in our analysis.}

\subsubsection*{Stability of the Potential} 

For the potential to be bounded from below (i.e., having a stable vacuum) the following conditions are required~\cite{Grzadkowski:2009bt}:
\bea
&& \lambda_{ii}>0 \quad (i =1,2,3),\label{positivity1} \\[2mm]
&& \lambda_x > - 2 \sqrt{\lambda_{11} \lambda_{22}}\,, 
\qquad \lambda_y > - 2 \sqrt{\lambda_{11} \lambda_{33}}\,, 
\qquad \lambda_z > - 2 \sqrt{\lambda_{22} \lambda_{33}}\,, \label{positivity2}\\[2mm]
&&\left\lbrace  \begin{array}{l} 
\sqrt{\lambda_{33}} \, \lambda_x + \sqrt{\lambda_{11}} \, \lambda_z+\sqrt{\lambda_{22}} \, \lambda_y \geq 0\\
\quad \textrm{or}\\[2mm]
\lambda_{33} \lambda_x^2 + \lambda_{11} \lambda_z^2+\lambda_{22} \lambda_y^2 -\lambda_{11} \lambda_{22} \lambda_{33} - 2 \lambda_x \lambda_y \lambda_z <0,
\end{array}\right.
\label{positivity3}
\eea
where 
\bea
\lambda_x = \lambda_{12}+\textrm{min}(0,\lambda_{12}'-2|\lambda_1|),\\[1mm]
\lambda_y = \lambda_{31}+\textrm{min}(0,\lambda_{31}'-2|\lambda_3|),\\[1mm]
\lambda_z = \lambda_{23}+\textrm{min}(0,\lambda_{23}'-2|\lambda_2|).
\eea
As noted in~\cite{Faro:2019vcd}, these conditions are in fact sufficient but
not necessary, as it is possible to construct examples of this model in which the potential is bounded from below, but which violate the conditions in Eqs.~\eqref{positivity1}--\eqref{positivity3}.  We do not explore such a region of parameter space in this work. 

\subsubsection*{Global Minimum Condition}

For $(0,0,v)$ to be a local minimum, all mass-squared values have to be positive and for it to be a global minimum, i.e., the true vacuum, its energy, $\mathcal{V}_{(0,0,v)}$, has to be lower than the energy of any other possible minima, $\mathcal{V}_{(v_i,v_j,v_k)}$, that exists at the same time. Following the chosen mass order and resulting relations in Eq.~\eqref{Hlightest}, we arrive at the following conditions:
\bea
 \textrm{local minimum if:}&& \left\lbrace \begin{array}{l}
v^2={\mu_3^2}/{\lambda_{33}} > 0,\\[1mm]
\Lambda_2 > \mu^2_2/v^2,  \\[1mm]
\Lambda_3 > \mu^2_1/v^2; \end{array} \right. 
\hspace{3cm}
\label{inert-loc}
\\
\textrm{global minimum if, in addition:}&& \mathcal{V}_{(0,0,v)}= -\frac{\mu_3^4}{4\lambda_{33}} < \mathcal{V}_{(v_i,v_j,v_k)}. 
\hspace{3cm}
\label{inert-glob}
\eea

\subsubsection*{Perturbative Unitarity} 

We require that the $2 \to 2$ scattering matrix is unitary, i.e., the absolute values of all eigenvalues of such a matrix for Goldstones, Higgs and dark states with specific hypercharge and isospin should be smaller than $8 \pi$. Furthermore, all quartic scalar couplings should be perturbative individually, i.e., { $|\lambda_i| \leq 4\pi$}.

\subsubsection*{EW Precision Observables (EWPOs)} 

We demand a 2$\sigma$, i.e., 95\% Confidence Level (CL), agreement with EWPOs which are  parametrised through the EW oblique parameters $S,T,U$. Assuming a SM Higgs boson mass of $m_h$ = 125 GeV, the central values of the oblique parameters are given by~\cite{Baak:2014ora}:
\be 
\hat{S} = 0.05 \pm 0.11 ,\qquad \hat{T} = 0.09 \pm 0.13, \qquad \hat{U}=0.01\pm 0.11.
\label{eq:ewpt}
\ee
In the I(1+1)HDM these constraints impose a strict order on the masses of the inert particles, with two neutral dark scalars being lighter than the charged particle. Furthermore, mass splitting between the heavier neutral scalar and the charged scalar is limited to roughly 50 GeV. However, in the case of a $\Z_2 \times \Z_2'$-symmetric I(2+1)HDM, these conclusions are no longer necessary. Cancellations between contributions to $S,T,U$ parameters from the two generations of dark particles may lead to { different mass orderings}, where either of $A_i$ or $H^\pm_i$ is the heaviest, as well as to an increased mass splitting between these particles (for a detailed discussion, see~\cite{Hernandez-Sanchez:2020aop}).

\subsubsection*{Collider Searches for New Physics} 

The presence of additional scalars, especially if they are sufficiently light, can influence properties of SM particles, e.g., their decay channels and widths. We forbid decays of EW gauge bosons into new scalars by enforcing:
\be 
\label{eq:gwgz}
m_{H_i}+m_{H^\pm_i}\,\geq\,m_W^\pm,~~~ m_{A_i}+m_{H^\pm_i}\,\geq\,m_W^\pm,~~~
\,m_{H_i}+m_{A_i}\,\geq\,m_Z,\,~~~
2\,m_{H_i^\pm}\,\geq\,m_Z.
\ee
Furthermore, we adopt LEP 2 searches for supersymmetric particles re-interpreted for the I(1+1)HDM in order to exclude the region of masses where the following conditions are simultaneously satisfied~\cite{Lundstrom:2008ai} ($i=1,2$):
\be 
\label{eq:leprec}
m_{A_i}\,\leq\,100\,\GeV,\,~~~
m_{H_i}\,\leq\,80\,\GeV,\,\, ~~~
\Delta m= |m_{A_i}-m_{H_i}|\,\geq\,8\,\GeV,
\ee
since this would lead to a visible di-jet or di-lepton signal.

The model also must agree with null results for additional neutral scalar searches at the LHC. 
As discussed in~\cite{Hernandez-Sanchez:2020aop}, current searches at the LHC for multi-lepton final states with missing transverse energy, $\met$, are, in general, not sensitive enough to probe the parameter space of our model. This is mainly due to a relatively large cut on $\met$ used in  experimental analyses, which results in a reduced sensitivity to probe the viable parameter space of the I(2+1)HDM scenario. Notice also that, as new charged particles are inert and hence do not couple to fermions, they are not subject to many constraints present in the 2HDM framework, e.g., flavour bounds on the charged scalar mass from $b\to s \gamma$ are not applicable here.

\subsubsection*{Charged Scalar Mass and Lifetime}

We take a model independent lower estimate on the masses of all charged states: $m_{H^\pm_i} > 70$ GeV ($i=1,2$)~\cite{Pierce:2007ut}. Furthermore, in this work we will not consider scenarios with possibly long-lived charged particles and, following~\cite{Heisig:2018kfq}, we set the limit for a charged state lifetime to be $\tau\,\leq\,10^{-7}\,{\rm s}$ .

\subsubsection*{Higgs Mass and Signal Strengths} 
The combined ATLAS and CMS result for the Higgs mass is~\cite{ATLAS:2015yey}:
\be
m_h = 125.09\pm 0.21 \textrm{ (stat.)} \pm 0.11 \textrm{ (syst.)} \; \GeV.
\ee
The Higgs particle detected at the LHC is in excellent agreement with the SM predictions. 
By construction, the $h$ state in the $(0,0,v)$ vacuum in Eq.~\eqref{eq:SM-higgs} is SM-like and its couplings to gluons, massive gauge bosons and fermions are identical to the SM values.
 
The Higgs total width can be modified through additional decays into light inert scalars, $S$, by contributing to the $h\to SS$ decay channel when $m_S \leq m_h/2$ as well as through modifications to decay channels already present in the SM, in particular, the $h \to \gamma \gamma$ and $\gamma Z$ decays. In this work we take the upper limit on the Higgs total decay width to be~\cite{Sirunyan:2019twz}:
\be
\Gamma_{\rm tot} \leq 9.1 \; \textrm{MeV}.
\ee
The partial decay width $\Gamma(h\to \gamma\gamma)$ is modified { with respect to the SM} through the presence of two charged inert scalars in the loop. In this work, we use the combined ATLAS and CMS limit for the signal strength~\cite{Khachatryan:2016vau}:
\be
\mu_{\gamma \gamma} = 1.14^{+0.19}_{{ -0.18}},
\ee
ensuring a 2$\sigma$ agreement with the observation. (Constraints from $\Gamma(h\to \gamma Z)$ are inferior in comparison.) 

{ The current 95\% CL upper limits on invisible Higgs-boson decays from CMS and ATLAS are~\cite{ATLAS:2023tkt,CMS:2023sdw}:}
\be
{ \textrm{BR}(h \to \textrm{inv.}) < 0.15 \; (\textrm{CMS}), \qquad \textrm{BR}(h \to \textrm{inv.}) < 0.107 \; (\textrm{ATLAS}).}
\ee
where we have introduced the Branching Ratio (BR).
These constraints significantly limit the allowed values of Higgs-inert state couplings for light inert states.

\subsubsection*{DM Constraints}
The total relic density is given by the sum of the contributions from both DM candidates $H_1$ and $H_2$,
\be 
\label{planck-relic}
\Omega_{\mathrm{DM}}h^2 = \Omega_{{H_1}}h^2 + \Omega_{{H_2}}h^2 \, ,
\ee
{ The observed abundance measured by Planck is}~\cite{Aghanim:2018eyx}:
\be
\Omega_{\text{\rm obs}}h^2 = 0.1200 \pm 0.0012.
\label{PLANCK_lim}
\ee
{ In our scan, this result is imposed as an upper bound on $\Omega_{\mathrm{DM}}h^2$, allowing the two scalar states to constitute only a fraction of the observed DM abundance. For an underabundant component $H_i$, we define $\xi_i=\Omega_{H_i}/\Omega_{\rm obs}$ and rescale the DD rate by $\xi_i$ and the ID annihilation rate by $\xi_i^2$.}
{ The strongest current limits on the Spin-Independent (SI) DM--nucleon scattering cross section, $\sigma_{\rm DM-N}$, in the mass range relevant to this study are provided by LUX-ZEPLIN (LZ), XENONnT and PandaX-4T~\cite{LZ:2024zvo,XENON:2025vwd,PandaX:2024qfu}.}
{ Regarding ID searches, for light DM particles annihilating into $b\bar b$ or $\tau^+\tau^-$, we apply the Fermi-LAT dwarf-spheroidal limits, which probe annihilation cross sections of order $3\times 10^{-26}~{\rm cm}^3{\rm s}^{-1}$ for $m_{\rm DM}\lesssim100~\GeV$~\cite{Ackermann:2015zua}. For heavier DM candidates, we use the antiproton and gamma-ray limits quoted in Ref.~\cite{Cirelli:2013hv}.}
\\



\section{Analysis Strategy at the LHC}
\label{sec:ana}

In this section, we describe in detail our numerical framework and collider analysis strategy.
The $\Z_2\times \Z_2^\prime$-symmetric 3HDM is implemented in \texttt{FeynRules}~\cite{Alloul:2013bka}, 
which is used to generate the model files both in UFO format and in the format required by 
\texttt{micrOMEGAs}~\cite{Alguero:2023zol}. Signal events are generated at Leading Order (LO) using \texttt{CalcHEP~3.9.2}~\cite{Belyaev:2012qa} and \texttt{MadGraph5\_aMC@NLO~3.5.6}~\cite{Alwall:2014hca} with the \texttt{NNPDF2.3LO} Parton Distribution Functions (PDFs), consistently with the event-generation description in Sec.~\ref{sec:bg}.
The subsequent parton showering and hadronisation are performed using \texttt{Pythia}~\cite{Bierlich:2022pfr}, 
while detector simulation is carried out with \texttt{Delphes}~\cite{deFavereau:2013fsa} within the \CM~\cite{Drees:2013wra,Kim:2015wza,Dercks:2016npn} framework, with resolution and ID parametrisation identical to the ATLAS 13 TeV parameters as implemented in \CM. 
The full cut-based analysis is also implemented within \CM.

\subsection*{Parameter Scan and Benchmark Selection}

A comprehensive scan over the model parameter space is performed to identify phenomenologically viable points satisfying all theoretical and experimental constraints discussed in Sec.~\ref{sec:cons}. 
The scan ranges are defined as follows:
\begin{align*}
&m_{H_1} \in [46.0,\,200.0]~\text{GeV}, ~~~~~~~~~~~~~~~~~~~~~~~~
m_{H_2} \in [46.0,\,200.0]~\text{GeV}, \\
&m_{A_1} \in [m_{H_1}+5.0,\, m_{H_1}+50.0]~\text{GeV}, ~~~~~~~~
m_{A_2} \in [m_{H_2}+5.0,\, m_{H_2}+50.0]~\text{GeV}, \\
&\Lambda_1 \in [-1.0,\,1.0], \qquad
\Lambda_2 \in [-1.0,\,1.0], \qquad
{ \Lambda_3 \in [-1.0,\,1.0].}
\end{align*}

The charged scalar masses are chosen sufficiently heavier than the corresponding neutral dark scalars in order to suppress co-annihilation effects. 
\begin{align*}
m_{H_1^\pm} = m_{H_1} + 50~\text{GeV}, \qquad
m_{H_2^\pm} = m_{H_2} + 50~\text{GeV}.
\end{align*}
The remaining quartic couplings in the inert sector ($\lambda_{11}$, $\lambda_{22}$, $\lambda_{12}$ and $\lambda'_{12}$) are fixed to $\sim 0.1$, 
as their variation has negligible impact on the collider phenomenology under consideration. 
For all parameter points, we ensure that $H_1$ and $H_2$ are the lightest neutral states in their respective $\Z_2$-symmetric sectors, thereby constituting two stable DM candidates.

After imposing theoretical consistency conditions (vacuum stability, perturbativity, and unitarity), collider constraints (Higgs signal strength and LEP/LHC bounds), and { DM constraints (the relic-density upper bound and the rescaled DD and ID  limits described in Sec.~\ref{sec:cons})}, we obtain a set of viable parameter points. From this set, we select representative BPs for detailed collider investigation.

\subsection*{Physics Motivation for the Benchmark Choice}

The primary objective of the collider analysis is to explore the distinctive phenomenological implications of having two DM candidates contributing simultaneously to the $\mu^+\mu^- + \met$ final state through the processes (see Fig.~\ref{fig:feyn})
\begin{align}
pp \to H_1 A_1 \to H_1 H_1 Z^*\to H_1 H_1 \mu^+\mu^-, \qquad
{ pp \to H_2 A_2 \to H_2 H_2 Z^*\to H_2 H_2 \mu^+\mu^-,}
\end{align}
alongside a jet from ISR.

To maximise the observability of a multi-component dark sector, we focus on scenarios where the two DM masses are sufficiently separated. In particular, we require
\begin{align*}
m_{H_1} \sim [50 - 75]~\text{GeV}, \qquad 
m_{H_2} \sim m_{H_1}+[30 - 50]~\text{GeV}.
\end{align*}
Additionally, the mass splittings with the corresponding CP-odd scalars are chosen to be 
\begin{align*}
\Delta m_1 = m_{A_1} - m_{H_1} \sim [35 - 55]~\text{GeV}, \qquad
\Delta m_2 = m_{A_2} - m_{H_2} \sim [15 - 25]~\text{GeV}.
\end{align*}

The kinematic structure of the signal is largely governed by the three-body decay $A_i \to H_i Z^* \to H_i \mu^+ \mu^-$. 
In the rest frame of $A_i$, the momentum of the dark scalar is given by
\begin{align}
p^*(H_i) = \frac{1}{2 m_{A_i}} 
\sqrt{\lambda(m_{A_i}^2, m_{H_i}^2, m_{\mu^+ \mu^-}^2)},
\end{align}
where $\lambda$ is the usual Källén function, usually defined as
\begin{align}
\lambda(x,y,z) = x^2 + y^2 + z^2 - 2xy - 2xz - 2yz,
\end{align}
see, e.g.,~\cite{Peskin:1995ev}.
{ The mass splitting also fixes the kinematic endpoint of the dimuon invariant-mass distribution, $m_{\mu^+\mu^-}^{\rm max}=m_{A_i}-m_{H_i}$. A larger splitting therefore produces a harder dimuon system with a higher endpoint, whereas a smaller splitting gives softer muons and a lower endpoint.}

{ The simultaneous contribution of two dark sectors with different values of $\Delta m_i$ can consequently produce a double structure in the dimuon invariant-mass spectrum. This is the characteristic feature that can distinguish the two-component scenario from a single-component inert scalar model. The hard ISR jet instead provides the recoil needed to generate large $\met$ and to trigger the event, i.e., it is not the origin of the double structure.}

\subsection*{Final Benchmark Selection: BP1}

Among the viable parameter points satisfying the above kinematic requirements, several candidates exhibit similar mass hierarchies. 
To select a single BP for detailed MC analysis, we adopt the following additional criteria.

\begin{itemize}
\item Maximisation of the combined production cross section 
      $\sigma(pp \to H_1 A_1 \to H_1 H_1 \mu^+ \mu^-) + \sigma(pp \to H_2 A_2 \to H_2 H_2 \mu^+ \mu^-)$.
\item Optimal signal efficiency after detector simulation and selection cuts.
\item Reduced overlap with dominant SM backgrounds in the $\mu^+\mu^- + \met +j$ channel.
\end{itemize}

\begin{table}[htbp]
        \centering
                \begin{tabular}{|c|c|c|c|c|c|c|c|c|}\hline
                        $m_{H_1}$&$m_{H_2}$&$m_{A_1}$&$m_{A_2}$& $m_{H_1^\pm}$&$m_{H_2^\pm}$&$\Lambda_1$ & $\Lambda_2 = \nicefrac{1}{2}g_{hH_2H_2} $ & $\Lambda_3 = \nicefrac{1}{2}g_{hH_1H_1}$   \\ \hline
                        57.92 & 93.98 & 103.19 & 115.7 & 107.92 & 143.9 & $-0.0026$ & 0.034 & 0.007
                        \\ \hline
		\end{tabular}
  \\[2mm]
   \begin{tabular}{|c|c|c|c|c|c|c|c|c|c|}\hline
                         $g_{hA_1A_1}$& $g_{hA_2A_2}$& $\Omega_{H_1} h^2$ &  $\Omega_{H_2} h^2$ & ${\rm BR}(h\to {\rm inv.})$&$\lambda_{11}$& $\lambda_{22}$ & $\lambda_{12}$ & $\lambda'_{12}$ &  $\lambda_{33}$  \\ \hline
                        0.255 & 0.22 & 0.003 & 0.0013 & 0.054 & $0.11$ & 0.12  & 0.121 & 0.13 & 0.129 
                        \\ \hline
		\end{tabular}
   \\[2mm]
   \begin{tabular}{|c|c|c|c|c|c|c|c|c|c|}\hline
                        $\lambda_1$&$\lambda_2$&$\lambda_3$&$\lambda_{31}$& $\lambda'_{31}$&$\lambda_{23}$& $\lambda'_{23}$ & $\mu^2_{1}$ & $\mu^2_{2}$ &  $\mu^2_{3}$  \\ \hline
                        { $-0.128$}  & { $-0.038$}  & $-0.06$  & 0.29  & $-0.15$  & 0.46  & { $-0.32$}  & $-2931.11$   & $-6774.7$ & 7812
                        \\ \hline
		\end{tabular}
        \caption{Details of BP1 in the $\Z_2\times\Z_2'$-symmetric I(2+1)HDM. Masses are given in GeV, and $v=246$ GeV. This BP passes all experimental and theoretical bounds, including { the relic-fraction-rescaled} ID and DD bounds on DM. (The cross sections are $\sigma_{\rm ID} = 8.6 \times 10^{-26}$ ${\rm cm}^3{\rm s}^{-1}$, ${\sigma_{\rm SI}^{\rm {H_1}}}=4.8 \times 10^{-10}$ ${\rm pb}$  and ${\sigma_{\rm SI}^{\rm {H_2}}}=4.6 \times 10^{-9}$~${\rm pb}$.) \label{tab:BP-ZZ-3}}
\end{table}

For each viable point, we compute the production cross sections at $\sqrt{s}=13.6$ TeV and fold them with detector-level efficiencies for signal selection. 
While multiple parameter points satisfy all theoretical and DM constraints and exhibit the desired mass hierarchy, the benchmark labelled \textbf{BP1} yields the largest effective signal rate after applying selection cuts. 
This enhancement arises from a favourable interplay between production cross sections, decay kinematics, and acceptance efficiencies, leading to improved signal-to-background separation relative to other candidate BPs.

The masses, relevant couplings, and DM properties of BP1 are summarised in Tab.~\ref{tab:BP-ZZ-3}. The benchmark considered here is underabundant and should be interpreted as a subdominant two-component scalar DM scenario. { Since the collider production and decay analysis is insensitive to the cosmological abundance, it remains applicable to models with the same electroweak production and cascade-decay topology.}
{ The region probed by the present LHC analysis requires relatively light electroweak states, ensuring sizeable $H_iA_i$ production cross sections, together with mass splittings large enough to yield reconstructable soft muons. In this region, the inert states also annihilate efficiently through electroweak gauge interactions. For BP1, $H_1$ lies close to the Higgs-resonance region, while $H_2$ is above the $W^+W^-$ threshold, further enhancing the annihilation rates and suppressing both relic-density components. Although the I(2+1)HDM can reproduce the full observed abundance, the corresponding regions generally involve masses and couplings less favourable for the present collider search. In addition, more compressed spectra reduce the acceptance for the soft muons and therefore further weaken the LHC sensitivity.}

In the following sections, we present a detailed collider analysis based on this BP to demonstrate { the LHC sensitivity to} the $\Z_2 \times \Z_2^\prime$-symmetric 3HDM in the $\mu^+\mu^- + \met +j$ channel.


\section{Background Estimation}
\label{sec:bg}
In our model, the $\mu^+\mu^- + \met \textcolor{blue}{+j}$
 signal, is
generated at the LHC by diagrams of the form given in Fig.~\ref{fig:feyn}, {where the radiated jet is omitted for clarity}.\footnote{In principle, 
$H_i^+H_i^-$ production processes can also contribute to the signal. However, the 
contributions from these processes were found to be sub-leading, at the level of less than 10\% of  $H_iA_i$ production $(i = 1,2)$.} The final state muons originate from the decay of the off-shell $Z$ boson for each DM generation.
{ As intimated, for each dark sector, the dimuon invariant-mass distribution has a kinematic endpoint at $m_{\mu^+\mu^-}^{\rm max}=m_{A_i}-m_{H_i}$, with $i=1,2$. The hard ISR jet enhances the signal $\met$ through recoil of the dark-sector system and, together with the large $\met$, provides the trigger handle: the analysis does not rely on triggering on the soft muons.}

The main LHC backgrounds for the signal under consideration come from { Drell--Yan production of $\mu^+\mu^-$ and $\tau^+\tau^-$ pairs through $Z^{(*)}/\gamma^{(*)}$,} SM production of $VV$ (where $V$ denotes the EW gauge bosons) and $Z\gamma^{*}$, and $t\bar t$ and $tW$ production
\cite{ATLAS:2017vat, ATLAS:2019lng}.
Additionally, a significant reducible background contribution arises from Fake Non-Prompt (FNP) leptons~\cite{ATLAS:2022swp}, which have been found substantial in experimental analyses. 
In general these can arise through three different physical processes: photon-to-lepton conversion, jet mis-identification, and heavy flavour, $b,c$-meson decays. However, if we limit ourselves to final-state muons the first two are negligible, making the FNP background less significant in the muon case (see Footnote 1). 
\begin{table}[htbp]
     \centering
           \begin{tabular}{c|c|c|c|c|c|c|c}\hline\hline
Process & $A_1H_1j+A_2H_2j$  &     $\mu^+\mu^-j$  & $Z^*/\gamma^*(\to{\tau^+\tau^-})j$ &  $W^+W^-j$ & { $Z\gamma^*j$} & $t\bar t+0,1j$ &$tW$\\ 
\hline
$\sigma$ [fb] &3.02+1.31  &  29,100  &  3,240   &  63.8  &  56.3 &  5,450  & 134\\
\hline\hline
\end{tabular}
\caption{Signal and background cross sections in ${\rm fb}$, at generator level.  All processes were independently generated at LO using {\madgraph} and \calchep, yielding consistent results. The cut $4\gev \lesssim m_{\mu^+\mu^-} \lesssim 60\gev$ was applied to these samples except $t\bar{t}$, while $p_T^j>50\gev$ was imposed where an additional jet was generated. The $t\bar{t}$ sample was merged and matched to one additional jet.
The signal is BP1 and the first (second) entry in the table is for $A_{1(2)}H_{1(2)}j\to \mu^+\mu^- H_{1(2)} H_{1(2)}j$, respectively. The backgrounds are:\,
        (1)~DY $\mu^+\mu^-j$;~
        (2)~$Z^*/\gamma^*j\to{\tau^+\tau^-j}\to\mu^+\mu^-j\nu_\mu\bar\nu_\mu\nu_\tau\bar\nu_\tau$;~
        (3)~$W^+W^-j\to \mu^+ \bar \nu_\mu\mu^-\nu_\mu~j$;~
        (4)~{ $Z(\to\nu\bar\nu)\gamma^*(\to\mu^+\mu^-)j$};~
        (5)~$t\bar{t} \to W^+W^-b\bar{b} \to \mu^+\mu^- \nu \bar\nu b\bar{b}$;~
        (6)~$tW\to t(t\to W^+ b\to b\mu^+\nu_\mu)W^-(\to \mu^-\bar\nu_\mu)$ and $\bar t(\bar t\to W^- \bar b\to \bar b\mu^-\bar\nu_\mu)W^+(\to \mu^+\nu_\mu)$.
        } \label{tab-ca-calchep}
\end{table}

All processes were independently generated at LO using {\madgraph}~3.5.6~\cite{Alwall:2014hca} and \calchep~3.9.2~\cite{Belyaev:2012qa}, yielding consistent results.  The samples from the former generator were used as the nominal samples for the analysis, with a generator-level requirement of $4\gev \lesssim m_{\mu^+\mu^-} \lesssim 60\gev$, and were subsequently showered with \pythia.
{ For samples containing an additional generated jet, a generator-level cut of $p_T^j>50\gev$ was imposed.} The semileptonic $t\bar{t}$ background was merged and matched up to 1 additional jet using \pythia.
For the $tW$ process generation we work in the 5-Flavour (5F) scheme.

We employ pre-selection requirements on the muons and jets
closely following Ref.~\cite{ATLAS:2017vat}. 
Preselected muons are required to pass the {\it Medium} criterion~\cite{ATLAS:2016lqx} with $p_T > 4 \gev$, $|\eta| < 2.5$.
Jets are reconstructed using the { anti-$k_T$} algorithm~\cite{Cacciari:2008gp, Cacciari:2011ma} with radius parameter $R = 0.4$
and are required to pass $p_T > 20 \gev, |\eta|<4.5$. The efficiency for $b$-jet identification
is assumed to be $85\%$.

Overlap removal between the muons and/or jets is ensured by imposing a set of 
conditions following Ref.~\cite{ATLAS:2017vat}. 
All non-$b$-tagged jets within
$\Delta R = 0.2$ of a pre-selected electron are removed. 
Here $\Delta R = \sqrt{(\Delta y)^2 + (\Delta \Phi)^2}$, $y$ being the rapidity and $\phi$ being the
azimuthal angle around the beam axis. Remaining jets containing less than two tracks with $p_T>0.5\gev$
within $\Delta R = 0.4$ of a pre-selected muon are removed. Electrons or muons with $\Delta R < 0.4$
from the remaining jets are then removed to suppress contributions from $b$- and $c$-hadron decays.

The FNP muon contribution to our signal region will predominantly be due to { weak decays of heavy hadrons}, primarily bottom and charm hadrons (see Fig.~5 of Ref.~\cite{ATLAS:2022swp}).  Since we are selecting events containing two opposite-sign (OS) muons, we would expect the dominant FNP muon contribution to stem from backgrounds with at least one prompt isolated muon, and one muon stemming from the decay of a bottom/charm hadron which is misidentified as a prompt muon.  The leading backgrounds would thus be $W(\to \mu\nu)+c$, $(W\to\mu\nu)+b\bar{b}$ and semi-muonic $t\bar{t}$ where both $b$-jets fail $b$-tagging. While FNP backgrounds are usually extracted using data-driven methods, { in this phenomenological projection we estimate them using simulated samples.}

Simulating the non-prompt muon background from heavy-flavour decays is non-trivial, both procedurally and from the point of view of statistics.  The production cross sections are large and the probability for the non-prompt muon to satisfy isolation criteria is dependent on precise details of showering, hadronisation and heavy-flavour decays, which are subject to significant modelling uncertainties.  Rather than expending significant computational and human resources simulating the `$W+$ heavy-flavour' background, we take advantage of the Open LHC Monte Carlo Event Generation datasets~\cite{ATLASOpenEventGen} made available by ATLAS as part of their open data offering.  We use `Specialised' dataset number \#501719, consisting of $W\to\mu\nu+0,\,1,\,2,\,3$~jets at { Next-to-LO (NLO)} for $\sqrt{s}=13$~TeV, with a multi-muon filter that captures heavy-flavour decays to muons.  The process has a cross section of 76.9~pb and sufficient statistics is  provided by ATLAS to reach an event weight of just over 2. { The event weights are rescaled to $\sqrt{s}=13.6$~TeV using the ratio of the corresponding inclusive $W+$jets cross sections at 13.6 and 13~TeV.} Note that parametric detector simulations like \texttt{Delphes} cannot simulate fake muons arising from interactions between light-flavour hadrons and detector materials, but this mechanism is not the leading source of FNP muons, as discussed above.
{ As shown in Tab.~\ref{tab:Cutflow}, after all cuts the FNP backgrounds remain subdominant to the total prompt SM background, although their modelling uncertainty must be included in a full experimental analysis.}  

\section{Results}
\label{sec:results}
We summarise our results in this section.
The leptons and jets satisfying the preselection and isolation conditions outlined in Sec.~\ref{sec:bg} are passed through the following
set of event selection criteria for our cut-and-count analysis.

\begin{enumerate}
\item
Exactly two tight OS muons  
.

\item 
A veto on $b$-tagged jets to suppress backgrounds from $t\bar t$ and $tW$ production.

\item 
$\met> 200~{\rm GeV}$ in order to reduce Drell-Yan and fake non-prompt backgrounds.

For the signal such a hard cut will require recoil of the parent system against a hard ISR jet (and is partially correlated with cut 5 below).

\item
$\Delta R (\mu^+, \mu^-) < 1.5$, consistent with a hard recoil.

\item
The leading jet must obey $p_T^{\rm j_1}>120 \gev$.

\item

$\Delta\phi(j_1,\met) > 2.0 $ to reduce the DY background where $\met$ arises due to jet mismeasurement.

\item 
$\Delta\phi(j,\met)>0.4$ for all jets.

\item
Reconstruct $m_{{\tau^+\tau^-}}$ by assuming energetic tau production in the $Z^{(*)}/\gamma^{(*)}(\to {\tau^+\tau^-})+ j$ process, i.e., assuming decay neutrinos are highly collinear with decay muons, and selecting
$m_{{\tau^+\tau^-}}<0$ or $m_{{\tau^+\tau^-}}>160~\text{GeV}$, where $m_{{\tau^+\tau^-}}$ is defined as $m_{\tau^+\tau^-} = \text{sign}\left(m_{\tau^+\tau^-}^{\rm 2}\right) \sqrt{|m_{\tau^+\tau^-}^{\rm 2}|}$~\cite{Han:2014kaa, Baer:2014kya, Barr:2015eva}.


\item 
$M_{T2}(\mu_1,\mu_2, \met)> 20\gev$ to further suppress the $Z^{(*)}/\gamma^{(*)}(\to {\tau^+\tau^-})$ background, for 
\begin{align}
M_{T2}(\mu_1,\mu_2,\met) = \min_{\vec{q}^{\,(1)}_T+\vec{q}^{\,(2)}_T={\cancel{\vec{p}}}_T} \left[\max\left[M_T(\vec{p}^{\,\mu_1}_T,\vec{q}^{\,(1)}_T),M_T(\vec{p}^{\,\mu_2}_T,\vec{q}^{\,(2)}_T)\right]\right],
\end{align}
with invisible test mass set to 0~\cite{Lester:1999tx,Barr:2003im,Cheng:2008hk}.

\item 
$M_{T2}(\mu_1\!+\!j,\mu_2\!+\!j,\met)>180\gev$ to reduce $t\bar t$ background events, for
{
\begin{align}
M_{T2}(\mu_1\!+\!j,\,\mu_2\!+\!j,\met) = \min\big[&M_{T2}(\mu_1+j_1,\mu_2+j_2,\met),\nonumber\\
&M_{T2}(\mu_1+j_2,\mu_2+j_1,\met)\big],
\end{align}
}
with invisible test mass set to 0.
\item
$m_{\mu^+\mu^-}<40\gev$ in order to avoid background  contributions from on-shell $Z$ boson decays.

\end{enumerate}

\begin{figure}[htb!]
       \vspace{1em}
\centering
\begin{subfigure}[b]{0.48\linewidth}
\centering\includegraphics[width=\textwidth]{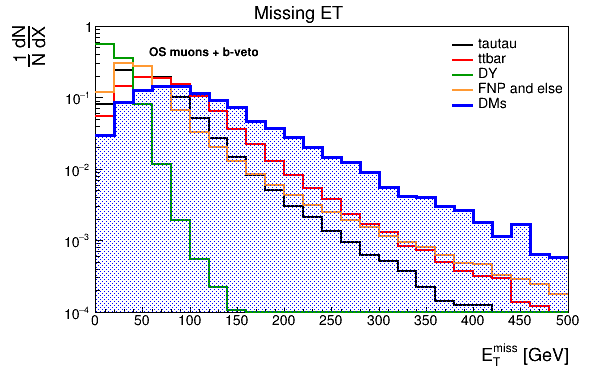}
        \caption{}
        \label{}
\end{subfigure}
~
\begin{subfigure}[b]{0.48\linewidth}
\centering\includegraphics[width=\textwidth]{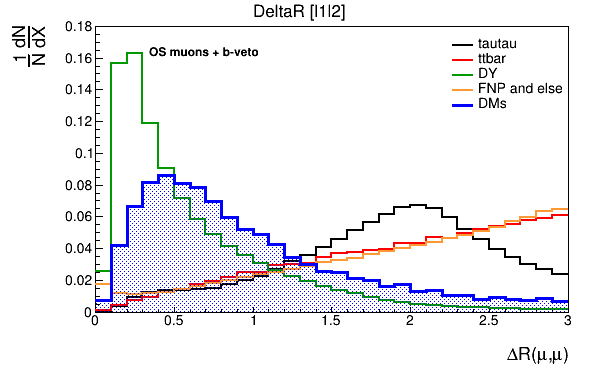}
        \caption{}
        \label{}
\end{subfigure}
\caption{The normalised distribution of the missing transverse energy $\met$ (left) and $\Delta R(\mu^-\mu^+)$ (right) after the pre-selection cuts and the requirement of OS muons and $b$-veto.}
\label{fig:fig_new-1}
\end{figure}

\begin{figure}[htb!]
       \vspace{1em}
\centering
\begin{subfigure}[b]{0.48\linewidth}
\centering\includegraphics[width=\textwidth]{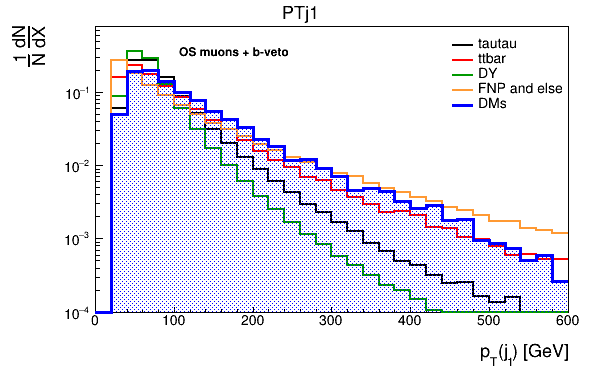}
        \caption{}
        \label{}
\end{subfigure}
~
\begin{subfigure}[b]{0.48\linewidth}
\centering\includegraphics[width=\textwidth]{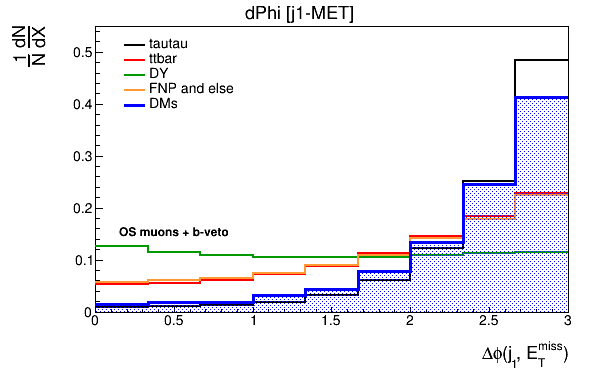}
        \caption{}
        \label{}
\end{subfigure}
\caption{The normalised distribution of the transverse momentum of the leading jet (left) and $\Delta \phi $ between the leading jet and $\met$ (right) after the pre-selection cuts and the requirement of OS muons and $b$-veto.}
\label{fig:fig_new-2}
\end{figure}

\begin{figure}[htb!]
       \vspace{1em}
\centering
\begin{subfigure}[b]{0.48\linewidth}
\centering\includegraphics[width=\textwidth]{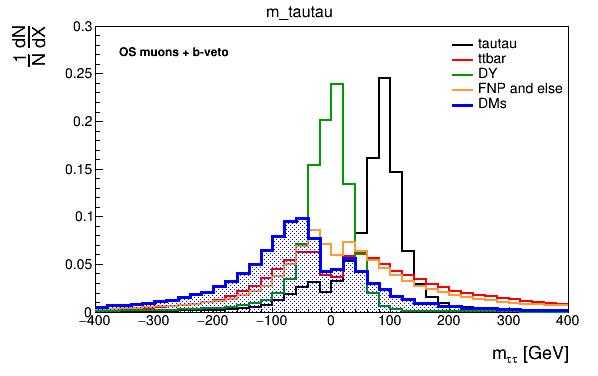}
        \caption{}
        \label{}
\end{subfigure}
~
\begin{subfigure}[b]{0.48\linewidth}
\centering\includegraphics[width=\textwidth]{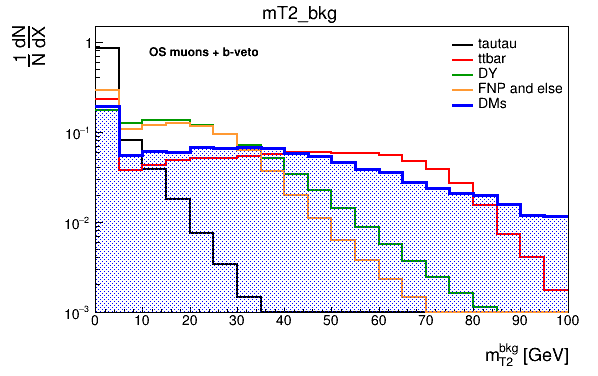}
        \caption{}
        \label{}
\end{subfigure}
\caption{The normalised distribution of the $m_{{\tau^+\tau^-}}$ (left) and $ M_{T2} (\mu^+,\mu^-,\met)$ (right) after the pre-selection cuts and the requirement of OS muons and $b$-veto.}
\label{fig:fig_new-3}
\end{figure}

\begin{figure}[htb!]
       \vspace{1em}
\centering
\begin{subfigure}[b]{0.48\linewidth}
\centering\includegraphics[width=\textwidth]{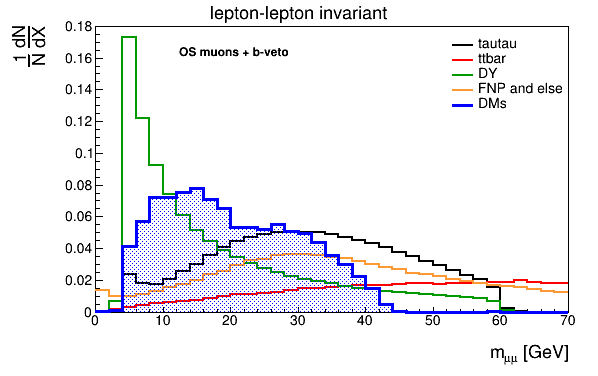}
        \caption{}
        \label{}
\end{subfigure}
~
\begin{subfigure}[b]{0.48\linewidth}
\centering\includegraphics[width=\textwidth]{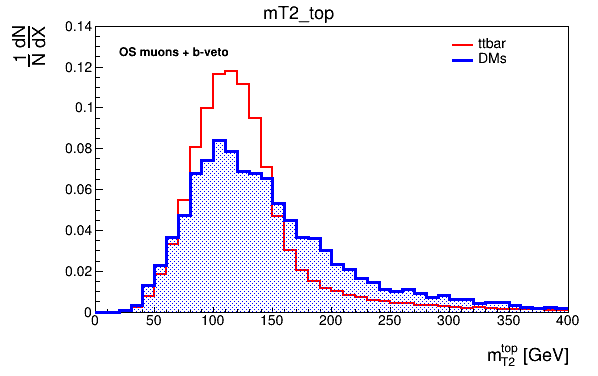}
        \caption{}
        \label{}
\end{subfigure}
\caption{The normalised distribution of the dimuon invariant mass (left) and $M_{T2}(\mu_1+j,\mu_2+j,\met)$ (right) after the pre-selection cuts and the requirement of OS muons and $b$-veto. { At this preselection level, note the double-bump structure in the dimuon invariant-mass distribution for the signal, the observation of which would be a smoking gun for the presence of two distinct DM sectors with independent masses and mass splittings.}}
\label{fig:fig_new-4}
\end{figure}

We show the normalised distributions for relevant collider observables at the LHC for the signal and background processes in Figs.~\ref{fig:fig_new-1}--\ref{fig:fig_new-4}. 
The distributions are shown for events satisfying the pre-selection 
criteria as well as the event selection cuts 1--2.
By analysing these distributions we further devise our advanced set of 
event selection cuts 3--11 to suppress the background to a manageable level and thus 
to achieve a viable signal significance at the LHC for our chosen
BP.  We would like to highlight our repurposing of the standard $M_{T2}^{\ell\ell}$ variable to exploit the difference between the topology of the dominant backgrounds and the unique topology of the signal, in which both muons arise from the decay of the same parent (see Fig.~\ref{fig:fig_new-3}).  A soft cut on this quantity (cut 9) is especially effective in reducing the ${\tau^+\tau^-}$ background, which decreases by a factor of just under 30.  { This motivates the relatively soft $M_{T2}^{\ell\ell}$ requirement used here, in contrast to the harder cuts commonly employed to suppress $W^+W^-$ and $t\bar t$ backgrounds in electroweakino searches with large chargino--neutralino mass differences~\cite{ATLAS:2019lng,ATLAS:2017vat,CMS:2021edw}.}

{In Fig.~\ref{fig:fig_new-1} we show the normalised distributions for the missing transverse energy $\met$ and radial angular separation between the two leptons in the final state after applying the pre-selection cuts and event selection cuts 1--2. Here, the contributions from the two DM candidates are summed over.
    Fig.~\ref{fig:fig_new-2} shows the normalised distributions for the $p_T$ 
    distributions of the leading jet and azimuthal angular separation between the leading jet and $\met$.
    Fig.~\ref{fig:fig_new-3} shows the normalised distributions for reconstructed $m_{\tau^+\tau^-}$ and $M_{T2}^{\ell\ell}$ ($\ell=\mu$).
    Fig.~\ref{fig:fig_new-4} shows the normalised distributions for the invariant mass of the final state OS di-muon system $m_{\mu^+\mu^-}$ and $M_{T2}(\mu_1\!+\!j,\mu_2\!+\!j,\met)$. The $m_{\mu^+\mu^-}$ distribution shows a kinematic endpoint at the larger of the two mass differences $m_{A_i}-m_{H_i}$, $i=1,2$. { At this preselection level, the signal also exhibits a double-bump structure characteristic of two distinct dark sectors with independent masses and mass splittings.}}


The cutflow corresponding to the selection cuts 
1--11 for our chosen BP is listed in Tab.~\ref{tab:Cutflow}
while the significance values for LHC Run 3 and HL-LHC are listed in Tab.~\ref{tab:significance_values}. For the nominal signal region, we obtain $S/B\simeq 9.8\%$ and $S/\sqrt{B}=1.35$ at $\mathcal{L}=300~\mathrm{fb}^{-1}$. A statistical-only extrapolation to $\mathcal{L}=4~\mathrm{ab}^{-1}$ gives $S/\sqrt{B}=4.93$, although a robust discovery-level conclusion would require a dedicated experimental treatment of the dominant background systematics.

{ Although the cut-based analysis provides sensitivity to the inclusive new-physics signal, it remains challenging to demonstrate that it originates from two DM candidates, since the double-bump structure is not retained with sufficient significance after the full selection. Employing a more sophisticated signal selection may increase our sensitivity to this characteristic feature, which we leave for future work.}

\begin{table}[htbp]
    \centering
    \resizebox{\textwidth}{!}{
    \begin{tabular}{c|c|c|c|c|c|c|c|c|c||c|c}
    \hline
        Cuts & DM1 & DM2 & Drell-Yan & $\tau^+\tau^-$ & dimuonic $t\bar t$ & $tW$ & $VV$ (all) & semi-muonic $t\bar t$ & $W+\mathrm{jets}$ & $S/B$ & $S/\sqrt{B}$ \\
        \hline \hline
        All & $1.81{\times}10^3$ & 786 & $8.73{\times}10^6$ & $9.73{\times}10^5$ & $1.64{\times}10^6$ & $1.60{\times}10^5$ & $9.34{\times}10^4$ & $9.79{\times}10^6$ & $2.41{\times}10^7$ & $5.7{\times}10^{-5}$ & 0.39 \\
        2 muons & 385 & 127 & $3.12{\times}10^6$ & $2.17{\times}10^5$ & $8.28{\times}10^5$ & $1.98{\times}10^4$ & $1.21{\times}10^4$ & $7.79{\times}10^4$ & $2.67{\times}10^6$ & $7.4{\times}10^{-5}$ & 0.19 \\
        OS tight muons & 385 & 127 & $3.12{\times}10^6$ & $2.17{\times}10^5$ & $8.26{\times}10^5$ & $1.97{\times}10^4$ & $1.20{\times}10^4$ & $4.21{\times}10^4$ & $1.83{\times}10^6$ & $8.4{\times}10^{-5}$ & 0.21 \\
        $b$-veto & 331 & 109 & $2.66{\times}10^6$ & $1.88{\times}10^5$ & $1.01{\times}10^5$ & $3.73{\times}10^3$ & $9.89{\times}10^3$ & $9.06{\times}10^3$ & $1.71{\times}10^6$ & $9.4{\times}10^{-5}$ & 0.20 \\
        $\met>200\gev$ & 34.1 & 16.1 & 149 & $1.88{\times}10^3$ & $2.68{\times}10^3$ & 102 & 388 & 197 & 823 & $8.1{\times}10^{-3}$ & 0.64 \\
        $\Delta R(\mu^+,\mu^-)<1.5$ & 29.5 & 14.8 & 145 & $1.79{\times}10^3$ & 523 & 58.1 & 251 & 41.4 & 200 & $1.5{\times}10^{-2}$ & 0.81 \\
        $p_T^{j_1}>120\gev$ & 24.6 & 13.6 & 102 & $1.75{\times}10^3$ & 486 & 55.6 & 233 & 39.4 & 186 & $1.3{\times}10^{-2}$ & 0.72 \\
        $\Delta\phi(j_1,\met)>2.0$ & 24.0 & 13.5 & 51.8 & $1.73{\times}10^3$ & 469 & 54.1 & 226 & 39.4 & 184 & $1.4{\times}10^{-2}$ & 0.71 \\
        $\Delta\phi(j,\met)>0.4$ & 21.8 & 12.3 & 25.0 & $1.62{\times}10^3$ & 318 & 45.4 & 207 & 29.6 & 161 & $1.4{\times}10^{-2}$ & 0.69 \\
        $m_{\tau^+\tau^-}>160\gev$ or $m_{\tau^+\tau^-}<0\gev$  & 18.6 & 10.6 & 22.7 & 290 & 277 & 41.0 & 180 & 27.6 & 148 & $3.0{\times}10^{-2}$ & 0.93 \\
        $M_{T2}(\mu_1,\mu_2,\met)>20\gev$ & 15.9 & 7.8 & 19.8 & 11.7 & 195 & 32.7 & 150 & 19.7 & 87.8 & $4.6{\times}10^{-2}$ & 1.04 \\
        $M_{T2}(\mu_1+j,\mu_2+j,\met)>180\gev$ & 13.1 & 5.8 & 13.4 & 8.2 & 74.6 & 20.8 & 116 & 7.9 & 40.1 & $6.8{\times}10^{-2}$ & 1.13 \\
        $m_{\mu^+\mu^-}<40\gev$ & $12.8$ & $5.8$ & $12.8$ & $5.1$ & $32.8$ & $12.9$ & $83.7$ & $5.9$ & $37.0$ & $9.8{\times}10^{-2}$ & 1.35 \\
        \hline \hline
    \end{tabular}
    }
    \caption{
    Cutflow for the signal and background event yields for LHC Run 3 at
    $\sqrt{s}=13.6~\mathrm{TeV}$ and $\mathcal{L}=300~\mathrm{fb}^{-1}$.
    }
    \label{tab:Cutflow}
\end{table}

\begin{table}[htbp]
    \centering
    
        \begin{tabular}{lcc}
        \toprule
         & LHC Run 3 $(300$ fb$^{-1})$ & HL-LHC $ (4$ ab$^{-1}) $ \\
         \midrule
        Signal events& $18.6 \pm 0.7$ & $248 \pm 9$\\
        Background events& $190.1 \pm 14.5$ & $2535 \pm 193$ \\ 
        \midrule
        $S/\sqrt{B}$ & 1.35 (1.91 naive combination) & 4.93 (6.97 naive combination)\\
        \bottomrule
    \end{tabular}
    
    \caption{{ Total signal and background yields and statistical-only significances for LHC Run 3 and the HL-LHC. The values in parentheses are naive ATLAS--CMS projections obtained by multiplying the single-experiment significance by $\sqrt{2}$, assuming identical sensitivities and statistically independent data sets: they do not represent an experimental combination.}}
    \label{tab:significance_values}
\end{table}

\begin{figure}[htb!]
\centering
\begin{subfigure}[b]{0.45\linewidth}
\centering\includegraphics[width=1.0\textwidth]{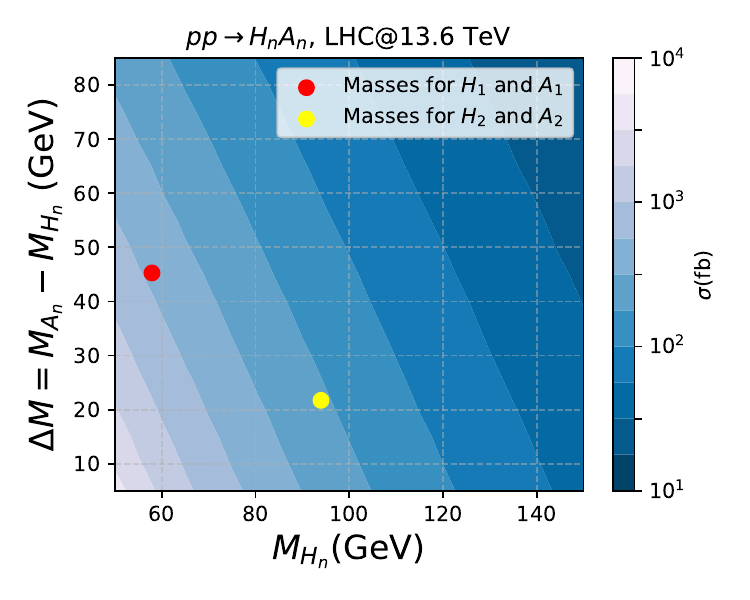}
	\caption{}
        \label{}
\end{subfigure}
~
\begin{subfigure}[b]{0.45\linewidth}
\centering\includegraphics[width=1.0\textwidth]{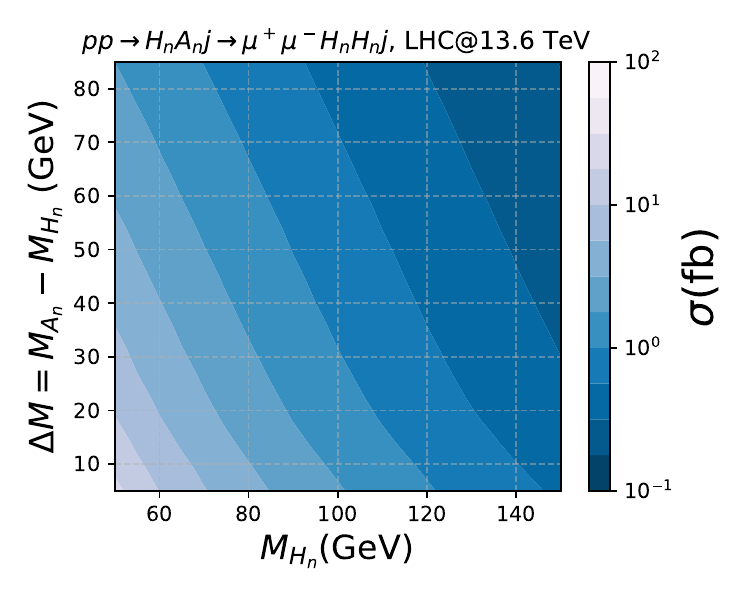}
	\caption{}
        \label{}
\end{subfigure}
~
\begin{subfigure}[b]{0.45\linewidth}
\centering\includegraphics[width=1.0\textwidth]{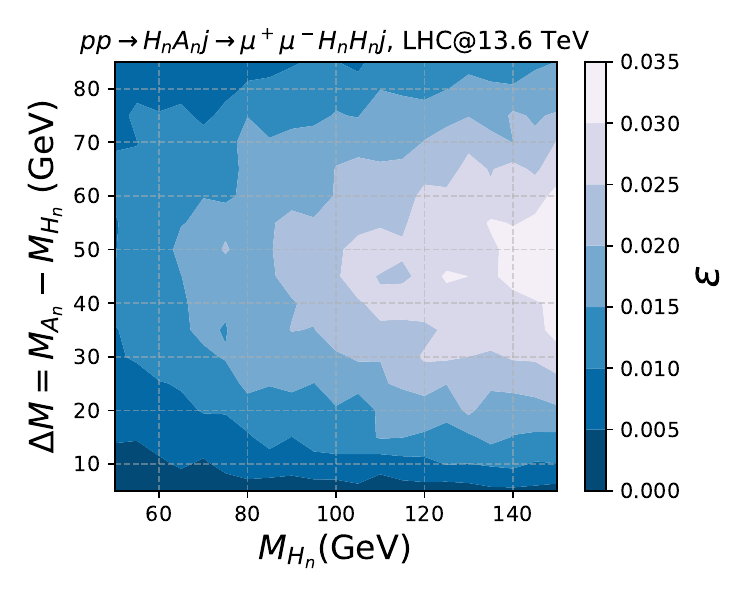}
        \caption{}
        \label{}
\end{subfigure}
~
%
\begin{subfigure}[b]{0.45\linewidth}
\centering\includegraphics[width=1.0\textwidth]{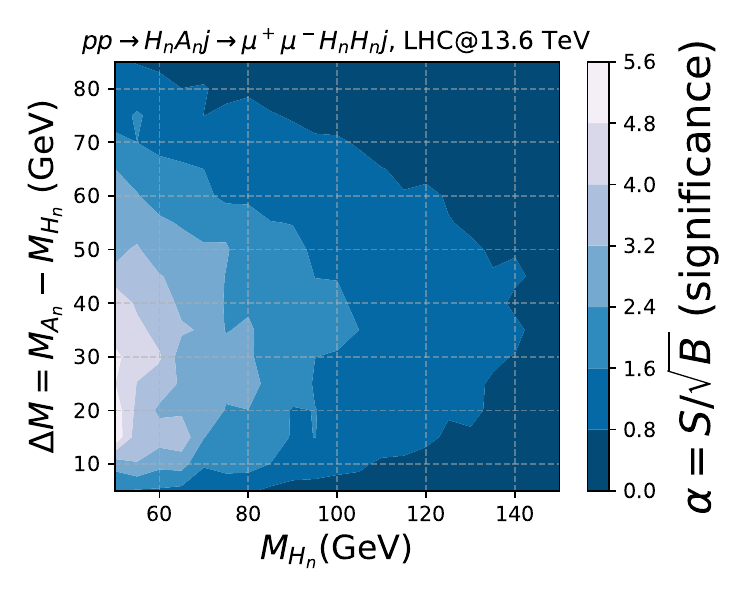}
\caption{}
\label{}
\end{subfigure}

\caption{Cross sections, efficiencies and significances at the LHC with $\sqrt{s}=13.6$ TeV over the I(2+1)HDM parameter space, mapped in terms of absolute mass of the DM candidate and the mass difference between this and the next-to-lightest neutral inert state
Subplot(a) also shows the explicit mass values chosen for BP1 for the first [red point] and second [yellow point] inert doublet. { Significance values in subplot (d) are given for the HL-LHC using a statistical-only extrapolation and are inclusive of both DM candidates.}}
\label{fig:scanplot}
\end{figure}

To give an overview of the
potential of the $\mu^+\mu^- + \met +j$ signal at the LHC to probe the parameter space
of our model beyond the BP we have discussed so far, 
we scan over $M_{H_1}$ and $\Delta M = M_{A_1}-M_{H_1}$, which are
the two parameters relevant for our kinematical analysis. Our results are depicted
in Fig.~\ref{fig:scanplot}, where we display the signal cross section $\sigma$ (top),
efficiency ($\epsilon = \frac{S}{\sigma.L}$) (bottom left) and the signal significance
$\alpha$ (bottom right) as functions of $M_{H_1}$ and $\Delta M = M_{A_1}-M_{H_1}$.
{ We note that the LEP measurement of the $Z$-boson width excludes the region $M_{H_1}+M_{A_1}<M_Z$~\cite{Cao:2007rm}, while the LEP-II reinterpretation excludes points for which $M_{A_1}\leq100\gev$, $M_{H_1}\leq80\gev$ and $\Delta M\geq8\gev$ are simultaneously satisfied~\cite{Lundstrom:2008ai}.} From the plots
it is clear that a statistical-only significance $S/\sqrt{B}\approx3$--$5$ at the HL-LHC
can be achieved in a significant region of the parameter space surviving the LEP constraints, { motivating dedicated searches for this topology at the HL-LHC.}

\clearpage
\section{Summary}
\label{sec:summary}
In this work, we explore a cut-based selection for soft muons arising from the production of two-component DM signals in the so-called I(2+1)HDM, i.e., a 3HDM with 2 inert and 1 Higgs doublets, { in the presence of} a $\Z_2\times \Z_2'$ symmetry. Our cut-based analysis shows that the benchmark BP1 can reach { $S/\sqrt{B}=1.35$ and $S/B\simeq 9.8\%$ at the LHC Run 3 with $300~\mathrm{fb}^{-1}$. A statistical-only extrapolation to $4~\mathrm{ab}^{-1}$ gives $S/\sqrt{B}=4.93$, indicating that the HL-LHC could reach discovery-level statistical sensitivity to this signal,} although a robust conclusion will require a dedicated treatment of experimental systematics. The use of muons, as opposed to also including electrons, { allows low transverse-momentum thresholds while reducing fake and non-prompt backgrounds.}

As part of our analysis, we repurpose a common $m_{T2}^{\ell\ell}$ cut to exploit the { signal} topology and reduce the $Z/\gamma\to\tau^+\tau^-$ background, one of the dominant backgrounds to this search, to manageable levels. Whilst a hard cut on this variable is commonly employed to minimise $WW$ and $t\bar{t}$ backgrounds in, e.g., electroweakino searches with large chargino-neutralino mass difference\footnote{Where each chargino decays by emitting an on-shell $W$.}, we give here a proof-of-principle example of the utility of this variable in a different context and suggest that it could enhance the sensitivity of other soft-lepton searches with a similar topology, e.g.~\cite{ATLAS:2019lng,ATLAS:2017vat,CMS:2021edw}.

{ The two dark sectors produce a characteristic double structure in the dimuon invariant-mass distribution before the full event selection. However, this feature is not retained with sufficient significance after the cuts optimised for inclusive signal sensitivity, making it difficult to establish experimentally that the signal originates from two DM components. More sophisticated analysis methods may improve the sensitivity to this characteristic structure.}

{ The benchmark considered here is underabundant and is therefore interpreted as a subdominant two-component DM scenario. Nevertheless, the collider analysis is independent of the cosmological abundance and motivates dedicated ATLAS and CMS searches for the soft-dimuon plus missing-energy signature. The results can also be reinterpreted in other weakly interacting DM scenarios with analogous electroweak associated production and cascade decays through an off-shell $Z$ boson.}

\subsection*{Acknowledgements}
We thank Tania Robens, with whom this project was initiated, and Krzystof Rolbieki for his prompt and frequent \checkmate\ support. AB and SM are supported in part through the NExT Institute and STFC CG ST/X000583/1. 
 AB acknowledges partial  support from Leverhulme Trust project MONDMag (RPG-2022-57).
 VK and AD acknowledge financial support from Research Ireland Grant 21/PATH-S/9475 (MOREHIGGS) under the SFI-IRC Pathway Programme. We acknowledge the use of the IRIDIS High Performance Computing Facility and associated support services at the University of Southampton in the completion of this work. All authors finally acknowledge financial support from the The Royal Society (London, UK) in the form of a Royal Society International Exchanges award, grant no. IES$\backslash$R1$\backslash$211138.  RM is supported in part by the Croatian Science Foundation (HRZZ) project “Beyond the Standard Model
discovery and Standard Model precision at LHC Run III”, IP-2022-10-2520.

\subsection*{Carbon Accounting}
We estimate the Green-House Gas (GHG) emissions of this work, taking into account plane travel between Zagreb and Southampton for the purposes of collaboration, calculated using the GES1point5 travel simulator tool~\cite{travelcalculator}, and event generation and analysis on the University of Southampton cluster.  For the latter we will use the methodology and formulas quoted in Ref.~\cite{LHCREIWGOpenEventGenerationTaskForce:2026dvu}, including the compute-per-event, and the processor specifications used therein. Since there is no public information on the PUE (a measure of the overhead energy use) of the Southampton data centre, we will use the industry average figure (for newer, energy-efficient data centres) of 1.3.  We use an average UK grid carbon intensity of 217 gCO$_{2e}$, taken from Ref.~\cite{OWID-gridcarbon}.  For the analysis, since the \checkmate\ tool works sequentially, we take the wall-clock time as a measure of compute. We count as negative emissions the GHG footprint avoided by not generating 10M $W+$heavy flavour events for our FNP background estimation, but instead downloading them from the ATLAS Open Event Generation dataset~\cite{ATLASOpenEventGen}. We assume that these events would otherwise have been generated on the Southampton cluster. The total GHG emissions of this work is approximately 5.2 tCO$_{2e}$, broken down as detailed in Tab.~\ref{tab:emissions}.

\begin{table}
    \centering
    \resizebox{0.6\textwidth}{!}{
    \begin{tabular}{l|l}
    \hline
    Source & Emissions [kgCO$_{2e}$] \\
    \hline \hline
    Plane travel & 4,900 \\
    30M generated events & 360 (use) + 99 (embedded) \\
    Analysis & 6.7 (use) + 1.6 (embedded) \\
    \textcolor{darkgreen}{Avoided generation of 10M events} & \textcolor{darkgreen}{- 120 (use) - 33 (embedded)} \\ 
    \hline
    \end{tabular}}
    \caption{Leading sources of GHG emissions for this work}
    \label{tab:emissions}
\end{table}

\bibliographystyle{unsrtnat} 
\bibliography{draft_i3hdm_2l.bib}

\end{document}